\documentclass[%
 reprint,
superscriptaddress,
showpacs,preprintnumbers,
 amsmath,amssymb,
 aps,
 pra
]{revtex4-1}

\usepackage{siunitx}
\usepackage{graphicx}
\usepackage{dcolumn}
\usepackage{bm}

\begin{document}


\title{Limitations of Particle-Based Spasers}

\author{G\"u{}nter Kewes}
\affiliation{AG Nanooptik, Institut f\"u{}r Physik, Humboldt-Universit\"a{}t zu Berlin, Newtonstrasse 15, 12489 Berlin, Germany}
\author{Kathrin Herrmann}
\affiliation{AG Theoretische Optik \& Photonik, Institut f\"u{}r Physik, Humboldt-Universit\"a{}t zu Berlin, Newtonstrasse 15, 12489 Berlin, Germany}
\author{Rogelio Rodr\'{i}guez-Oliveros}
\affiliation{AG Theoretische Optik \& Photonik, Institut f\"u{}r Physik, Humboldt-Universit\"a{}t zu Berlin, Newtonstrasse 15, 12489 Berlin, Germany}
\author{Alexander Kuhlicke}
\affiliation{AG Nanooptik, Institut f\"u{}r Physik, Humboldt-Universit\"a{}t zu Berlin, Newtonstrasse 15, 12489 Berlin, Germany}
\author{Oliver Benson}
\affiliation{AG Nanooptik, Institut f\"u{}r Physik, Humboldt-Universit\"a{}t zu Berlin, Newtonstrasse 15, 12489 Berlin, Germany}
\author{Kurt Busch}
\affiliation{AG Theoretische Optik \& Photonik, Institut f\"u{}r Physik, Humboldt-Universit\"a{}t zu Berlin, Newtonstrasse 15, 12489 Berlin, Germany}
\affiliation{Max-Born Institut, Max-Born-Strasse 2a, 12489 Berlin, Germany }

\date{\today}

\begin{abstract}
We present a semi-classical analytic model for spherical core-shell surface plasmon lasers. 
Within this model, we drop the widely used one-mode approximations in favor of fully electromagnetic Mie theory. This allows for incorporation of realistic gain relaxation rates that so far have been massively underestimated. Especially, higher order modes can undermine and even reverse the beneficial effects of the strong Purcell effect in such systems. Our model gives a clear view on gain- and resonator-requirements, as well as on the output characteristics that will help experimenters to design more efficient particle-based spasers. 
\end{abstract}

\maketitle


Nanoscopic sources of coherent electromagnetic fields are essential elements for different fields in nanooptics, such as nanoplasmonics \cite{Stockman2011a}, metamaterials \cite{Hess2012b}, and quantum plasmonics \cite{Tame2013}. A surface plasmon laser (spaser) might be such a nanoscopic source \cite{Bergman2003,Ma2013a}. 

As compared to a laser, the obvious difference of a spaser is the use of plasmons instead of photons. Plasmons are inherently localized excitations and generally exhibit much smaller (mode) volumes than photonic cavity modes \cite{Koenderink2010a}. From an electromagnetic perspective, there is no reason to expect any further principle deviations from well-known (semi-classical) laser physics. For instance, Mie theory \cite{Bohren2008} completely describes the electromagnetic field for a spherical particle irrespective of the constituent material, i.e., whispering gallery modes of dielectric spheres and localized plasmon modes of metallic spheres are all included. However, as we will detail below, there are certain issues that have to be treated with care.  

Recently, a number of spaser devices\,\cite{Noginov2009,Suh2012,Meng2013a,Oulton2009,Ma2011,Khajavikhan2012,Lu2012a} have been characterized and extensive theoretical work has addressed fundamental and device-specific spaser properties\,\cite{Khurgin2012, Andrianov2011, Andrianov2011a, Parfenyev2012, Ginzburg2013, Stockman2010}. However, several questions, even of a fundamental nature, remain to be answered. Perhaps the most important of these is related to the rather low spaser efficiency. For instance, previous experiments placed very high demands on the pump (e.g., high laser pulse intensities \cite{Meng2013a}), the synthesis of the spaser's gain medium (e.g., dense incorporation of fluorophores \cite{Noginov2009}), and, quite generally, very high demands regarding the material quality \cite{Lu2012a}. Accordingly, these issues are reflected by the rather small number of publications that address spaser action in fully nanoscopic systems and systems working with organic gain media\,\cite{Noginov2009,Suh2012,Meng2013a}. 
 
Analytic theoretical descriptions have mainly focused on quasi-static analysis, so far\,\cite{Bergman2003,Andrianov2011,Andrianov2011a, Parfenyev2012, Ginzburg2013, Stockman2010}. Within this framework only a nanoparticle's dipolar resonance or generic numbers of the gain medium's relaxation rate have been considered to describe the spaser. Rather numerical cold cavity analysis of actual devices has been used in order to show (i) correspondence with observed far-field patterns and measured spectra etc.\ \cite{Oulton2009,Khajavikhan2012,Ma2011,Lu2012a}, (ii) that the resonator under observation is unable to support ordinary purely optical modes \cite{Oulton2009,Khajavikhan2012,Ma2011,Lu2012a}, and (iii) to estimate the minimum gain required to overcome losses within the resonator. Within these studies the gain-factor was basically described by a constant negative imaginary part of the dielectric permittivity.

A complete description of spasing/lasing systems requires a fully quantum-mechanical treatment. In view of the dispersive and dissipative properties of metals and the open-system character of any spaser, this represents a formidable task. In this work, we set ourselves the more modest, but also challenging goal of developing a fully electromagnetic semi-classical rate-equation approach. This allows for a quantitative investigation of the stationary and time-dependent (see supplemental material (SM)) input-output characteristics, the necessary inversion density (or gain-factor) and unmasks the so far disregarded impact of higher order modes. 

The paper is organized as follows: We start with a description of the model and how we apply it to the spaser design published by Noginov et al.\,\cite{Noginov2009}. The results found for this design are presented and discussed before we conclude.

\begin{figure}
\includegraphics *[width=0.99\columnwidth]{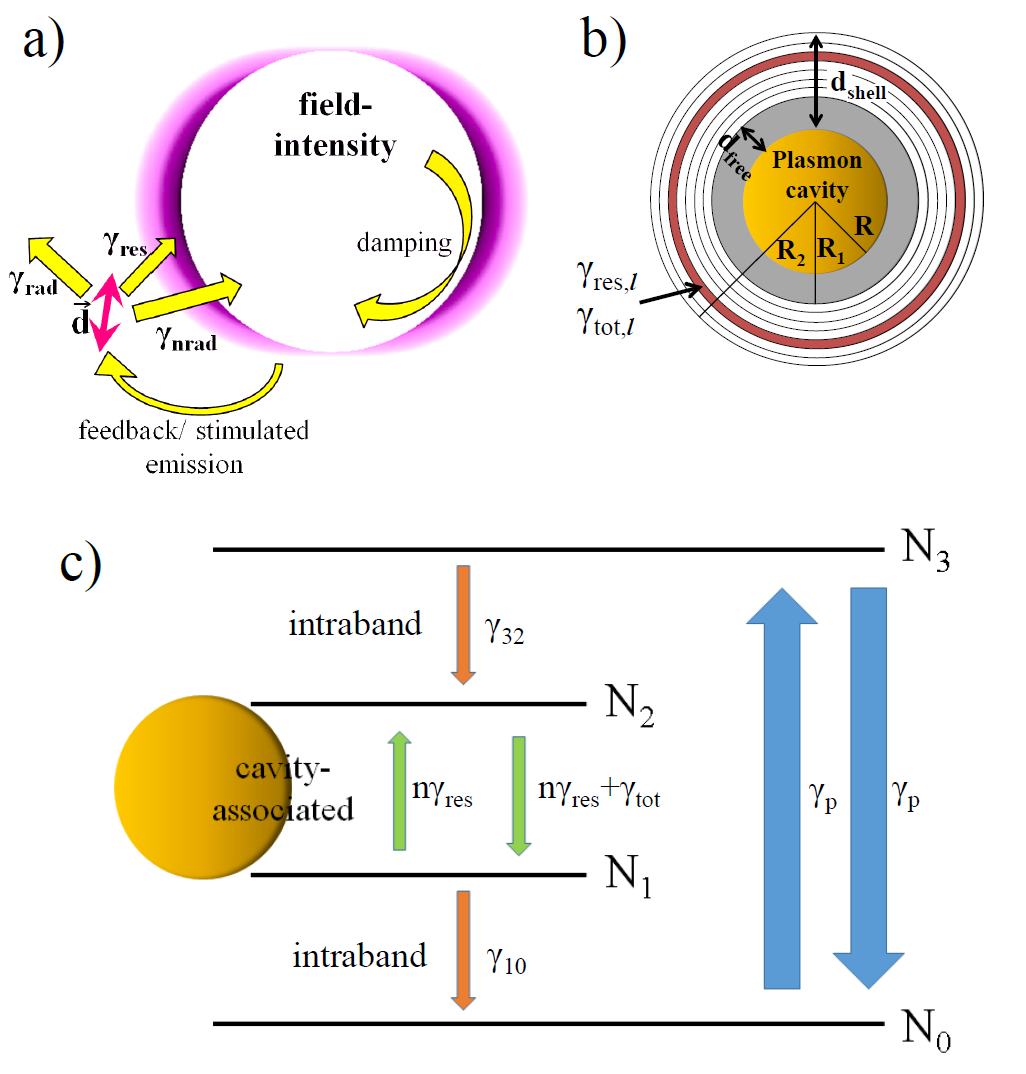}
\caption{\label{fig1}Physical model of gain medium and loss processes for a spaser that operates with emitters and the dipolar resonances of a metal sphere. 
(a) The gain-medium model considers dipolar emitters ($\vec{d}$) in proximity to a lossy plasmonic resonator, specifically the decay rates into different channels like off-resonant higher order modes ($\gamma_{\rm{nrad}}$), the far-field ($\gamma_{\rm{rad}}$) and the resonator-mode ($\gamma_{\rm{res}}$). 
(b) The emitters are randomly but homogeneously incorporated in a shell around the sphere of thickness $d_{\rm{shell}}$ (between the radii $R$ and $R_{2}$) except for the emitter-free spacing layer (gray) of thickness $d_{\rm{free}}$ ($R$ to $R_{1}$). The part of the shell that hosts emitters is treated as a series of layers (with index $l$ and accordingly with individual decay rates $\gamma_{\rm{res,}\textit{l}}$ and $\gamma_{\rm{tot,}\textit{l}}$) to account for radial changes in the inversion. 
(c) Scheme of the four-level gain medium taken as a basis for the rate-equations. The intraband relaxation rates $\gamma_{32}$ and $\gamma_{10}$ are assumed to be of equal speed, for simplicity: $\gamma_{\rm{intra}}$=$\gamma_{32}$=$\gamma_{10}$. $\gamma_{\rm{tot}}$ denotes the total decay rate and $\gamma_p$ the pump rate}
\end{figure}

We consider a spherical metal resonator coated by a shell of gain material. The description of the resonant mode in our model is based on the analytically derived eigenfrequency and quality factor of a dipolar mode of a metal sphere \cite{Kolwas2006}. The gain medium is described by an ensemble of four-level emitters interacting with all the different multi-polar 'modes' supported by the resonator (Fig.\ \ref{fig1}). We are starting from the well-known decay properties of a single emitter placed in front of a spherical particle \cite{Ruppin1982a} and construct a rate-equation description for an arbitrary number of homogeneously distributed and randomly oriented emitters that constitute the gain medium. This multi-emitter system is then merged into an approximate system that depends on averaged emitter properties only. To still account for radially variable gain depletion the averaging is executed within thin sub-layers of the gain medium shell. In a last step the rate-equation system is solved.

The accurate description of relaxation rates inside the gain-medium is an essential part of our model. The spontaneous decay rates of \textit{individual} emitters in the vicinity of spherical metal particles have been obtained some time ago using a multi-polar (Mie) expansion of the electromagnetic field \cite{Ruppin1982a}. Ruppin's framework has influenced several work on single emitter emission \cite{Chew1987,Kim1988,Carminati2006,ColasdesFrancs2008,Moroz2010} and  is also the basis of our gain-medium description. In particular, Ruppin found that for dipole emitters in close proximity (sub-wavelength distance) to the metal sphere, coupling of the emitter to quadrupolar and higher-order multi-poles contributes significantly to the spontaneous decay rate even when the emitter is resonant with the dipolar mode. The excitation of off-resonant higher order plasmonic 'modes' equates to non-radiative decay channels. This is sometimes referred to as surface energy transfer (SET) \cite{Yun2005,ColasdesFrancs2008} or more general as quenching.
This is also related to the fact that for dispersive and dissipative material properties, the multi-polar 'modes' are not power orthogonal. 
Physically speaking, these multi-polar 'modes' exchange power between each other and, considering that higher-order multi-poles do not radiate but are prone to Ohmic losses, this leads to additional non-radiative decay processes. In fact, such cross-coupling effects are well-known in laser physics (excess noise or Petermann factor) \cite{Siegman1989} and lead to modifications of the laser threshold and laser linewidth. Obviously, such effects cannot be captured when only considering dipolar modes or generic numbers for the relaxation rate of the gain medium \cite{Bergman2003,Andrianov2011,Andrianov2011a, Parfenyev2012, Ginzburg2013, Stockman2010}. 
Specifically, the decomposition of the rates into multi-polar orders (Mie coefficients) allows us further to analyze in detail where the emitted energy goes: to Ohmic losses/dissipation ($\gamma_{\rm{nrad}}$), to far-field radiation ($\gamma_{\rm{rad}}$), or to the actual resonator-mode ($\gamma_{\rm{res}}$). Further, one can derive directly from these calculations relevant numbers such as the $\beta$-factor $\beta=\gamma_{\rm{res}}/\gamma_{\rm{tot}}$ (with $\gamma_{\rm{tot}}$=$\gamma_{\rm{nrad}}$+$\gamma_{\rm{rad}}$) or the Purcell-factor $\Gamma=\gamma_{\rm{res}}/\gamma_0$. Note, that also $\gamma_{\rm{res}}$ is partly radiative and non-radiative. Here, $\gamma_0$ is the unperturbed vacuum decay rate of the emitter that is needed as an experimental input to calculate the absolute values of the various decay rates. Since the decay rates  $\gamma(\vec{r},\vec{p})$ of individual emitters are extremely sensitive to the precise position $\vec{r}$ and orientation $\vec{p}$ of the emitter \cite{Ruppin1982a,ColasdesFrancs2008}), the representation of the gain medium by an \textit{ensemble} of dipolar emitters with \textit{different} decay properties is a key feature of our model. This clearly goes beyond the usual linear gain medium description, especially as spatial hole burning effects can be included. Details about the calculation of the decay rates and the rate-equation description of the spaser system are given in the SM. 

In the following we consider a spaser that operates on the (three energetically degenerate) dipolar resonances of a gold sphere. We refer to this system as a lasing spaser, since the dipolar mode is 'bright', i.e., this mode  couples to the far-field and thus emits photons. 
We utilize the spaser geometry as reported by Noginov et al.\,\cite{Noginov2009}: The gold plasmon resonator of radius $R$=\SI{7}{\nano\meter}
is centered in a spherical shell of thickness $d_{\rm{shell}}$=\SI{15}{\nano\meter} and refractive index $n_{\rm{out}}$=1.46 that is doped with organic emitters (OG-488). For these parameters we find a complex dipolar eigenfrequency of $\omega_{\rm{res}}$=$\SI{2.29-i0.123}{\eV}$ which yields a quality factor of $Q_{\rm{res}}$=$\rm{Re}[\omega_{\rm{res}}]/(-2\,\rm{Im}[\omega_{\rm{res}}])$=$9.29$ (Noginov et al.\,found $Q_{\rm{res}}$=14.8). This corresponds to a loss-rate of the resonator of 2\,$\rm{Im}[\omega_{\rm{res}}]$=$\SI{5.96e13}{\s^{-1}}$. To calculate the absolute values of the decay rates of the emitters we use the measured lifetime of $\tau$=\SI{4.3}{\nano\second} in ethanol \cite{Noginov2009}. With a refractive index of ethanol $n_{\rm{eth}}$=1.33 we extract the corresponding vacuum decay rate of $\gamma_{0}$=1/($n_{\rm{eth}}\tau)=\SI{1.75e8}{\s^{-1}}$. For the intraband relaxation inside the 4-level gain medium we chose an optimistically high value of $\gamma_{\rm{intra}}$=$\SI{e13}{\s}^{-1}$ (typical values range from $\SI{e11}{\s}^{-1}$ to $\SI{e13}{\s}^{-1}$ \cite{Sugawara1997}). Throughout the paper gold is modeled via a Drude-Lorentz permittivity that was fitted to the data measured by Johnson $\&$ Christy \cite{Johnson1972a} (fitting parameters see SM). \par

\begin{figure}
\includegraphics *[width=0.99\columnwidth]{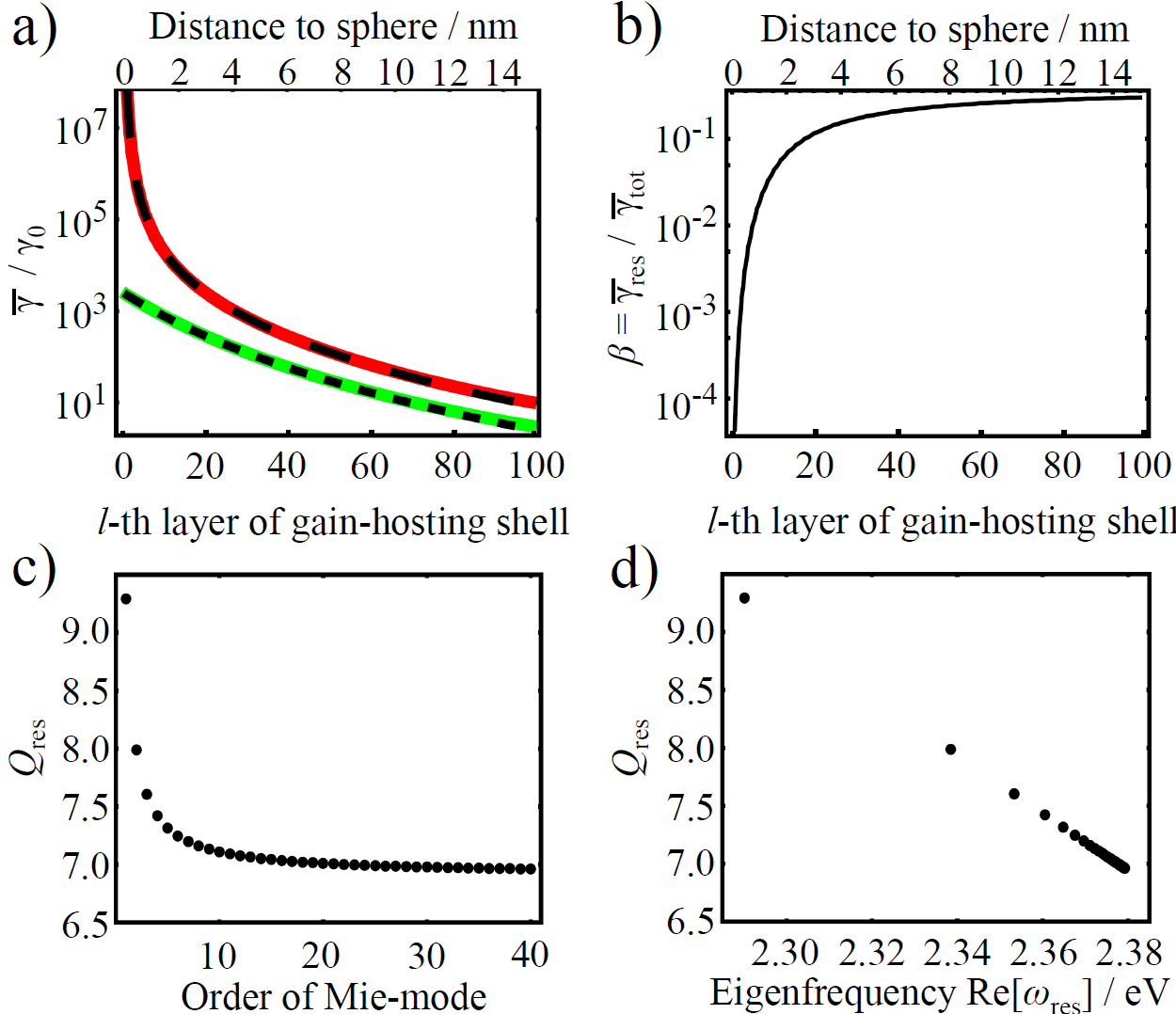}
\caption{\label{fig2}Angular- and orientation-averaged decay rates of a dipolar emitter as a function of layer $l$ inside the shell (equivalent to distance to the sphere) and quality factors $Q_{\rm{res}}$ of higher order modes in the gold nanosphere ($R$=\SI{7}{nm} and outer refractive index $n_{\rm{out}}$=1.46). The emission frequency of the emitter is resonant with the sphere's dipole modes ($\lambda_{\rm{res}}$=\SI{537}{\nano\meter}). All rates are normalized to the corresponding vacuum decay rate $\gamma_0$. 
(a) The red curve shows the (relative) total decay rate $\bar{\gamma}_{\rm{tot}}/\gamma_0$. The sparsely dashed black curve shows the non-radiative decay rate $\bar{\gamma}_{\rm{nrad}}/\gamma_0$ that almost perfectly overlaps with $\bar{\gamma}_{\rm{tot}}/\gamma_0$. The green curve shows the decay rate into a single spaser mode $\Gamma$=$\bar{\gamma}_{\rm{res}}/\gamma_0$, i.e., the Purcell-factor. The tighter dashed black line shows the non-radiative part of $\Gamma$, showing that also $\Gamma$ is almost exclusively non-radiative. 
(b) Ratio of the decay rate into the spaser mode to the total decay rate $\bar{\gamma}_{\rm{res}}$/$\bar{\gamma}_{\rm{tot}}$, i.e., the $\beta$-factor. (c) Quality factors of the first 40 multi-polar Mie-modes. Higher order modes have $Q_{\rm{res}}$ comparable with that of the dipolar resonance. (d) $Q_{\rm{res}}$ as a function of frequency. The resonances lie very close to each other, i.e., they 'condensate' \cite{Kolwas2006,Andrianov2014}.}
\end{figure}

First, we take a look at the decay properties of the emitters embedded in the gain medium shell. As mentioned above and described in detail in the SM, the decay rates ($\gamma_{\rm{tot}}(\vec{r},\vec{p})$ and $\gamma_{\rm{res,}}(\vec{r},\vec{p})$) enter our model in an averaged manner ($\bar{\gamma}_{\rm{tot,}\textit{l}}$ and $\bar{\gamma}_{\rm{res,}\textit{l}}$), where the average is taken over possible positions and orientations within thin sub-layers $l$ of the gain medium shell. To obtain a reasonable resolution we consider 100 sub-layers. As shown in Figure \ref{fig2}a) and b), an average emitter located in the outermost sub-layer $l=100$  at distance of $\approx\SI{15}{\nano\meter}$ from the sphere experiences a relatively high $\beta$-factor ($\beta$=$\bar{\gamma}_{\rm{res,100}}$/$\bar{\gamma}_{\rm{tot,100}}$) of roughly $\SI{30}{\%}$, however only a moderate Purcell-factor $\Gamma$=$\bar{\gamma}_{\rm{res,100}}/\gamma_0\approx3$ with an actual coupling rate of $\bar{\gamma}_{\rm{res,100}}\approx$ $\SI{5.25e8}{\s^{-1}}$ is obtained. This value is especially small compared to the loss-rate of the spaser resonator ($\SI{5.96e13}{\s^{-1}}$). For emitters closer to the metal sphere the coupling to the spaser mode increases by orders of magnitude up to $\bar{\gamma}_{\rm{res,1}}\approx2000{\cdot}\gamma_0=\SI{3.5e10}{\s^{-1}}$ for the innermost sub-layer $l$=$1$. However this comes at the price of an over-proportionally enhanced total decay rate of $\bar{\gamma}_{\rm{tot,1}}\approx$ $\SI{7e15}{\s^{-1}}$, due to the non-radiative coupling to higher order modes and consequently a poor $\beta$-factor of around $10^{-4}$. This demonstrates that $\bar{\gamma}_{\rm{res,}\textit{l}}$ may become orders of magnitudes smaller than $\bar{\gamma}_{\rm{tot,}\textit{l}}$ and thus close-by emitters can only provide a very small ratio of their emitted energy to the actual spaser mode. Consequently, there are two reasons for which the feeding of the spaser's resonator mode may become inefficient: The low coupling to the resonator mode for emitters at the outer edge of the shell and the high coupling to off-resonant modes for emitters close to the metal sphere. The latter effect is well known (since it limits performance) in the field of plasmonic nanoantennas where often one is interested in boosting the radiative decay of single photon emitters \cite{Schietinger2009}. So far, this effect has largely been ignored in spaser theories although coupling to higher order modes is non-negligible already in the quasi-static description \cite{Moroz2010}. However, while negligence of off-resonant modes is often justified in ordinary laser systems, it has no base when emitters are close to a metal sphere. Especially for small spheres, where the quasi-static limit is accurate, the quality factors of higher order modes are comparable to that of the dipolar mode (Fig.\ \ref{fig2}(c)). Furthermore, quite counter-intuitively these resonances 'condensate' \cite{Kolwas2006,Andrianov2014} at a close-by frequency (Fig.\ \ref{fig2}(d)), thus overlapping with the emission spectrum of the gain-medium. In fact, we summed up the contributions of the first 400 multipolar orders to obtain a reasonably accurate total decay rate for the innermost layer $\bar{\gamma}_{\rm{tot,1}}$ (see also \cite{Ruppin1982a,Moroz2010}).\par
Before we solve the spaser rate-equation we perform a simple estimate for the minimal emitter number that is needed to sustain spasing. The resonator losses need to be compensated by stimulated processes in the gain medium. Assuming a complete inversion this yields:\newline
\begin{equation}
\frac{N \overline{\gamma_{\rm{res}}}}{2|\rm{Im}[\omega_{\rm{res}}]|}\geq 1 \\
\label{eq2}
\end{equation}
where, $N$ is the total number of emitters in the shell, $\overline{\gamma_{\rm{res}}}$ is the decay rate of the emitters into the active mode averaged over the entire shell and $\rm{Im}[\omega_{\rm{res}}]$ is the imaginary part of the complex eigenfrequency of the dipolar mode. For the described spaser design we obtain a minimal emitter number $N_{\rm{min}}$=$4518$, which corresponds to an emitter density of $\rho_{\rm{min}}$=$\SI{1.05e20}{cm^{-3}}$, similar to the emitter number reported by Noginov et al.\,\cite{Noginov2009}. With the emission (absorption) cross-section $\sigma_{\rm{em}}$\,$\approx$\,$\sigma_{\rm{abs}}{=}\SI{2.55e-16}{cm^{-2}}$\,\cite{Noginov2009} we obtain the corresponding minimal gain-factor of  $g$=$\rho_{\rm{min}}{\cdot}\sigma_{\rm{em}}$=$\SI{26775}{cm^{-1}}$, which is already extremely high, especially for organic gain media (see following discussion). Note, that this minimal gain estimate is in good consistency with general electrodynamic predictions based on the quasi-static approximation \cite{Wang2006,Arnold2015}. Solving the rate-equations we will however see, that it is almost impossible to reach complete inversion and thus even higher emitter densities are needed.

\begin{figure}
\includegraphics *[width=0.99\columnwidth]{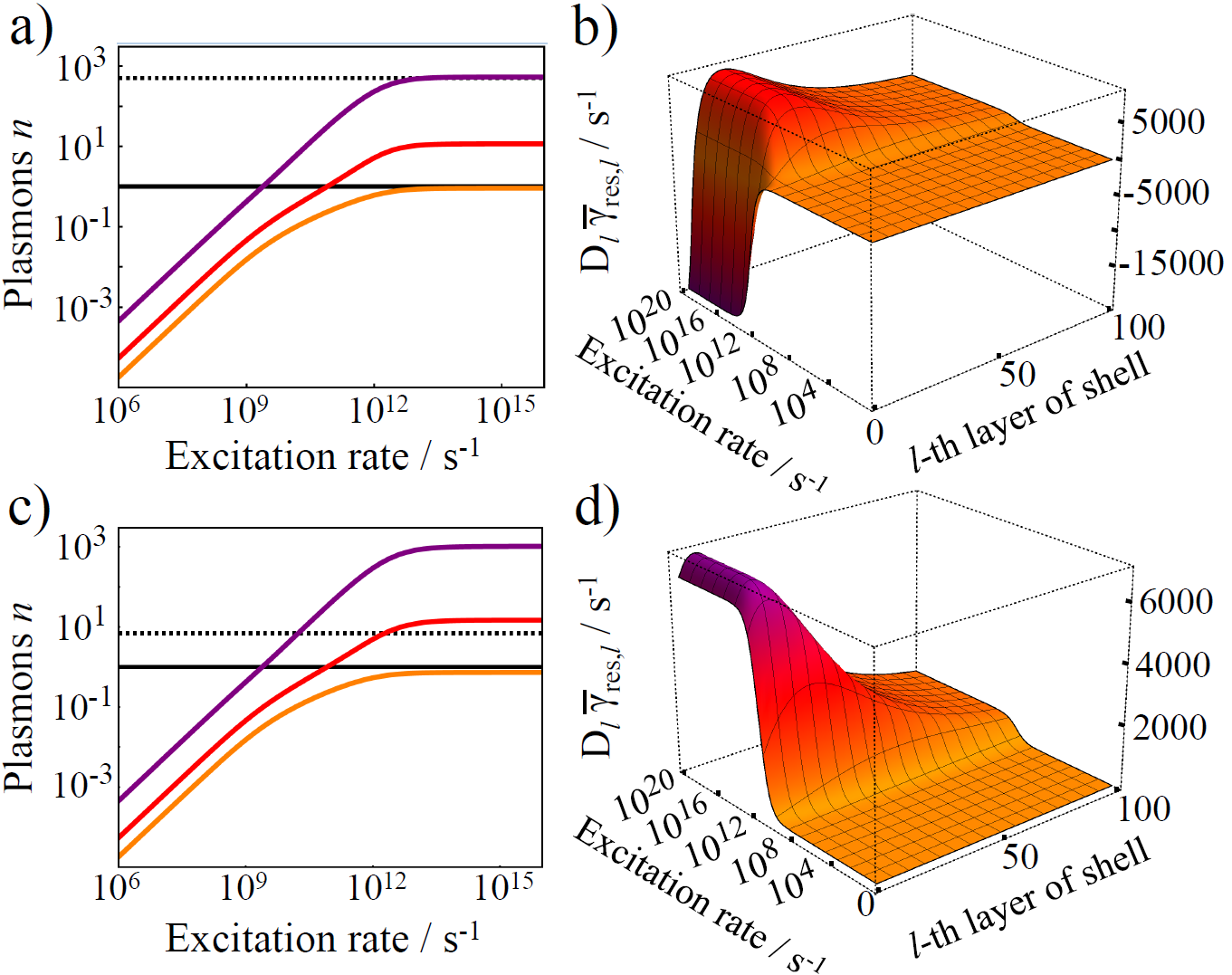}
\caption{\label{fig3}Stationary solution of the rate-equations: (a,b) without ($d_{\rm{free}}$=$\SI{0.1}{\nano\meter}$) and, (c,d) with a $\SI{2}{\nano\meter}$ spacing layer between the emitters and the sphere surface. (a,c) Plasmon number $n$ in the resonator-mode of a core-shell particle as a function of pump rate per emitter (double-logarithmic plot) for three different emitter-densities in the shell corresponding to 1$\cdot \rho_{\rm{min}}$, 3$\cdot \rho_{\rm{min}}$ and 25$\cdot \rho_{\rm{min}}$ (orange, red and purple line, respectively). The horizontal lines mark the threshold population numbers (solid: $n$=$1$, dashed: $n$=$\bar{\beta}^{-1}$ with $\bar{\beta}^{-1}\,\approx\,500$ in a) and $\bar{\beta}^{-1}\,\approx\,7$ in c)) (b,d) Rate of absorbed or gained plasmons ($D_l \bar{\gamma}_{\rm{res,\textit{l}}}$) per layer $l$ as a function of pump-rate corresponding to the red curves with 3$\cdot \rho_{\rm{min}}$ in (a,b).}
\end{figure}

Further, as elaborated earlier by others, the condition of loss-compensation by stimulated emission does not represent a rigorous definition of a laser's threshold (as it can never be fulfilled in a stationary solution) \cite{Bjoerk1994,Rice1994}. A common definition is given by the following condition: to have an equal amount of spontaneous and stimulated relaxation processes that populate the resonator mode. When considering only the resonant mode this is equivalent to an average population with a single photon/plasmon of ($N_2\overline{\gamma_{\rm{res}}}n=N_2 \overline{\gamma_{\rm{res}}}\Leftrightarrow n$=$1$) \cite{Bjoerk1994}. However, in the case of a particle-based spaser two distinct zones are of special interest: near- and far-field. Initially, the spaser was proposed to be a source of coherent near-fields \cite{Bergman2003}. Especially here the first definition of $n$=$1$ falls short, as the near-field will be occupied  by a significant amount of (excess) noise due to the open-system character and the population of higher order modes. Thus, the near-field will be highly incoherent at this low population. Taking into account the spontaneous emission into higher
order modes, the mentioned condition (spontaneous=stimulated emission) is $N_2\overline{\gamma_{\rm{res}}}n=N_2 \overline{\gamma_{\rm{tot}}}$ which yields $n$=$\bar{\beta}^{-1}$ needed plasmons in the spaser mode. Further, due to saturation effects in the gain-medium, the population of the partly radiative dipolar mode lacks an obvious kink in the input-output curve (Fig. \ref{fig3}), which would be a typical indication for the onset of lasing. Consequently, an experimental verification of a threshold by analyzing the light emitted to the far-field will be extremely difficult. 

At this point we are ready to compute the actual stationary input-output curves of the spaser. Fig. \ref{fig3}a) shows the number of plasmons in the resonator-mode as a function of pump rate (per emitter) for various emitter densities. 
The horizontal lines mark the aforementioned threshold populations of $n$=$1$ and $n$=$\bar{\beta}^{-1}$ where we have averaged the $\beta$-factor over the entire shell. In Fig. \ref{fig3}b) we plot the weighted inversion $D_l \bar{\gamma}_{\rm{res,\textit{l}}}
$ for each $l$, which is the rate of absorbed (for negative values) or gained plasmons in the various sub-layers. 

Using the above estimated minimal emitter density $\rho_{\rm{min}}\approx\SI{1.05e20}{cm^{-3}}$ (orange line) we observe that no spasing is established as the average plasmon number stays below the $n$=$1$ level. The level of $n$=$\bar{\beta}^{-1}$ (here $\bar{\beta}^{-1}$$\approx\,500$) is only crossed when extreme emitter densities are assumed that are roughly 25-times higher (purple line). This result can be explained when looking at the weighted inversion in Fig. \ref{fig3}b) (plotted for $\rho$=$3{\cdot}\rho_{\rm{min}}$). At low pump powers we find an inversion of $D_l\approx 0$ for all  sub-layers. For layers at distances of $\gtrsim$\,$\SI{2}{nm}$ from the core's surface the inversion grows at higher pump-rates. However in layers closer to the gold core, we find an inversion that finally becomes negative. This part of the shell becomes absorptive, thus \textit{preventing} spaser action. 

Now we incorporate an emitter-free spacing layer of thickness $d_{\rm{free}}$ between gold core and doped shell, where the emitter density and the outer shell radius are unchanged with regard to the previous design. Fig. \ref{fig3}c,d) shows the results for $d_{\rm{free}}$=$\SI{2}{nm}$. In this case a positive inversion is maintained and way less noise is present due to the better average $\beta$-factor of about $\bar{\beta}$$\approx$$14\,\%$. Thus spasing in the sense of a coherent near-field ($n\approx\bar{\beta}^{-1}\approx7$) is reached for significantly lower emitter densities of 3$\cdot \rho_{\rm{min}}$. However, this emitter density and especially the corresponding gain-factor is still extremely high.

The results found in the last section are surprising at first glance: First, the gain-factor corresponding to the 4518 emitters is already very high but even stronger gain-media are needed. Second, the gain-medium in the most intensive zone of the plasmon field close to the gold core (experiencing the highest Purcell-factor) is actually disadvantageous for spaser action.
Emitters close to the resonator produce a lot of noise and efficiently absorb emerging coherent fields and funnel the energy into lossy channels (higher order modes). The absorptive nature emerges even in a four-level gain medium since the induced decay rates are so fast, that they can keep up with the intraband relaxation inside the gain medium. The lower level of the optically active transition is not emptied fast enough, so that it acts as a blocked drain and charge carriers accumulate that lead to the aforementioned re-absorption. In literature, such effects are known as 'phonon bottleneck', but so far they only occurred for gain media with slow intraband relaxation rates as known from quantum dots \cite{Sugawara1997}. In contrast here we assumed fast relaxation rates as known for quantum wells or bulk. The slower the intraband relaxation rates the wider the zone of negative inversion around the gold core will be. A way to counter this issue is obviously to introduce a protective emitter-free spacing layer between metal and gain medium as shown in Fig. \ref{fig3}.

Our conclusion to use emitter-free spacing layers is in very good agreement with results from larger spaser devices that do not utilize localized but propagating plasmons and semiconductors as gain medium \cite{Oulton2009,Khajavikhan2012,Ma2011,Lu2012a}. These designs make use of a 'hybrid' propagating mode that is guided mainly inside a spacing layer between high-index semiconductor gain-materials and a metal film to minimize propagation losses \cite{Oulton2008}. As we now indirectly demonstrated, these designs are avoiding the problematic issue with enhanced loss channels.

We have mentioned, though, that even with a spacing layer extremely high inversion densities (or gain-factors) are needed to achieve spasing. The simple estimation for $\rho_{\rm{min}}$ and the associated gain-factor $g$=$\rho_{\rm{min}}{\cdot}\sigma_{\rm{em}}$ assuming full inversion falls short because of strong and in-homogeneous gain depletion. The results do of course depend on the precise spaser design and the intraband relaxation rates. Note, that the model assumes that the emitters are uncoupled and their properties are maintained irrespective of their dense packing. However, gain-media composed of densely packed organic dye molecules run into limitations quite early (formation of non-fluorescent complexes and other quenching mechanisms) \cite{Gather2014}. Due to such effects, realistic gain-factors are typically limited to not more than $\SI{100}{\cm^{-1}}$ \cite{Chenais2011}. The operation of a particle-based spaser with organic dye molecules, therefore, seems unrealistic on the whole and especially for the design of Noginov et al.\,\cite{Noginov2009}. 

Finally, relevant for experimenters, the input-output curves reveal, that the population of the resonator with plasmons will saturate relatively fast (Fig.\ref{fig3}) in the range of 10-100 plasmons as the cycle of charge carriers is limited by the intraband relaxation rates. Thus the near-field will neither show a high degree of coherence nor the extremely high intensity predicted by earlier studies. Further, only a small percentage of the plasmons will leave the resonator to be detected in the far field as even the so-called 'bright' dipolar mode, is almost exclusively fed by non-radiative processes as can be seen from Fig. \ref{fig2}a).

To conclude, we introduced a spaser model that explicitly includes quenching processes due to higher order modes and applied it to the experimental work of Noginov et al \cite{Noginov2009}. We found surprisingly pessimistic results that are caused by: \\
(i) the low $Q$-factor of the resonator, i.e., high loss-rates that require very high gain-factors for compensation; \\ 
(ii) strong quenching close to the resonator, which undermines (reverses) potentially beneficial effects expected from high Purcell-factors. Quenching comes along with poor $\beta$-factors which is the source for a significant amount of noise in the spaser's near-field. Further, the fast quenching rates destroy the inversion as these processes can keep up with the intraband relaxation rates.

How to tackle these issues? Low $Q$-factors and strong quenching are directly connected to the imaginary part of the dielectric permittivity $\epsilon''$ of the gold resonator. To reduce these material-related losses one needs to use silver instead of gold or to shift the working frequency of the spaser towards the red or near infrared by either using elongated rods or by embedding spheres into higher refractive index materials. However, quenching will remain strong close to the resonator, thus a spacing layer between resonator and gain-medium is inevitable. Finally, semiconductor gain-media are needed, which can provide significantly higher gain-factors of more than $\SI{1000}{\cm^{-1}}$ and do not suffer from bleaching. These insights should help experimenters to design particle-based spasers with significantly improved efficiencies and to estimate their performance. The results are further relevant for the metamaterials community. The insights can be actually used one-to-one, when it comes to attempts to compensate for losses in metamaterials with the help of gain-media.
Our model can be suitably enhanced by numerical approaches \cite{Sauvan2013a,Agio2013a,Kristensen2014} for determining 'quasi-normal modes' for complex plasmonic nanoparticles. So, it provides a rather easy tool to use for computing minimum requirements for a given design and will thus be extremely helpful for future experimental works. In addition, our semi-classical model may provide the basis for full quantum-optical treatment of spaser 
action.

We acknowledge support by the Deutsche Forschungsgemeinschaft (DFG) through the sub-projects B2 and B10 within the Collaborative Research Center (CRC) 951 Hybrid Inorganic/Organic Systems for Opto-Electronics (HIOS). We thank Prof.\;Thomas Klar for his valuable comments and fruitful discussions. G.K.\;and K.H. contributed equally to this work.

\end{document}